\documentstyle[prl,twocolumn,aps,floats,epsfig]{revtex}

\begin{document}

\draft
\twocolumn[\hsize\textwidth\columnwidth\hsize\csname
@twocolumnfalse\endcsname

\title{The oxygen isotope effect on critical temperature in superconducting 
copper oxides} 

\author{A. Mourachkine} 

\address{Nanoscience, University of Cambridge, 11 J. J. Thomson Avenue, 
Cambridge CB3 0FF, UK} 

\date{Received 10 December 2003}
\maketitle

\begin{abstract} 

The isotope effect provided a crucial key to the development of the BCS 
(Bardeen-Cooper-Schrieffer) microscopic theory of superconductivity for 
conventional superconductors. In superconducting cooper oxides (cuprates) 
showing an unconventional type of superconductivity, the oxygen isotope effect 
is very peculiar: the exponential coefficient strongly depends on doping level. 
No consensus has been reached so far on the origin of the isotope effect in 
the cuprates. Here we show that the oxygen isotope effect in cuprates is 
in agreement with the bisoliton theory of superconductivity.  

\end{abstract}

\pacs{PACS numbers: 74.62.-c; 74.62.Dh; 74.72.-h} 
]

Just before the development of the BCS theory (Bardeen-Cooper-Schrieffer) 
\cite{BCS}, it was experimentally found that different isotopes of mercury 
have different critical temperatures, and $T_c M^{1/2} =$ {\em constant}, 
where $M$ is the isotope mass \cite{6,7}. In the framework of the BCS theory, 
the expression for the critical temperature $T_c$ is given by 
$k_BT_c = 1.14\, \hbar \omega _D \exp ( - 1/\lambda)$, where $\omega _D$ 
is the phonon frequency at the edge of the Debye sphere; $\lambda$ 
is the electron-phonon coupling constant (= $N(0)V$); $k_B$ is the Boltzmann 
constant, and $\hbar$ is the Planck constant divided by $2\pi$. So, the BCS 
theory has provided an explanation for the isotope effect because the Debye 
frequency varies as $\omega _D \propto M^{-1/2}$, and consequently, the product 
$T_c M^{1/2}$ is constant for a given superconductor. For the majority of 
superconducting elements, the exponential coefficient $\alpha$ is indeed close to 
the classical value of 1/2. Later on, it was discovered that the situation is more 
complicated than it had appeared to be. For some conventional superconductors, 
the exponent 
of $M$ is not 1/2, but near zero. For example in Ru and Zr, $\alpha = 0 \,\pm$\,0.05. 
Thus, the isotope effect is not a universal phenomenon, and can be absent even 
in {\em conventional} superconductors. Nevertheless, this effect played 
the decisive role in showing the way to the correct theory of superconductivity 
in metals. 

In superconducting cooper oxides (cuprates) \cite{Muller}, the situation is very 
peculiar. The oxygen isotope effect was found to be extremely small in 
optimally doped cuprates. This fact was initially taken as evidence against the BCS 
mechanism of high-$T_c$ superconductivity ({\em true}) and, mistakenly, against 
the phonon-pairing mechanism ({\em false}). If the pairing mechanism is different 
from the BCS mechanism, {\em this does not mean that phonons are irrelevant}. 

In fact, there is a huge isotope effect in the cuprates. Figure 1 
shows the oxygen ($^{16}$O vs $^{18}$O) isotope-effect coefficient 
$\alpha _O = {\rm d} (\ln T_c)/{\rm d} (\ln M)$, where $M$ is the isotope mass, 
as a function of doping level in La$_{2-x}$Sr$_{x}$CuO$_{4}$ (LSCO), 
YBa$_{2}$Cu$_{3}$O$_{6+x}$ (YBCO) and Bi$_{2}$Sr$_{2}$CaCu$_{2}$O$_{8}$ 
(Bi2212). In the plot, one can see that the oxygen-isotope effect in the 
cuprates is not universal: it is system- and doping-dependent. In the 
underdoped region, $\alpha _O$ can be much larger than the BCS value of 0.5 
(according to the BCS theory, the isotope effect cannot be larger than 0.5). 
In the optimally doped region ($p \sim$ 0.16), the oxygen-isotope effect is 
indeed small. With the exception of one point in LSCO, $p =$ 0.125 (the so-called 
$\frac{1}{8}$ anomaly in LSCO), the doping dependence of the coefficient $\alpha_0$ 
is universal in these cuprates. The coefficient $\alpha _0$ has a maximum at 
$p \rightarrow$ 0.05. 

\begin{figure}[t]
\epsfxsize=0.8\columnwidth
\centerline{\epsffile{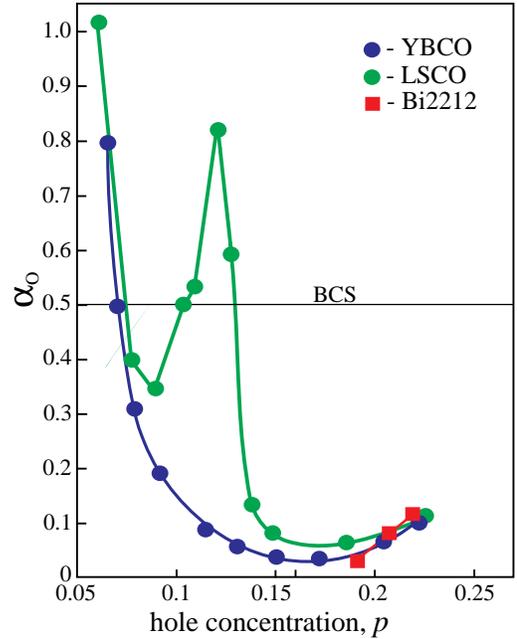}}
\caption{Oxygen isotope effect ($^{16}$O vs $^{18}$O) in LSCO, YBCO and Bi2212 
(taken from Ref. \protect\cite{R}).}
\end{figure} 

Even if a few attempts have been made in the literature to explain 
such a peculiar isotope effect in the cuprates \cite{isotope}, no consensus 
has been reached so far. The main 
purpose of this paper is to show that the isotope effect in cuprates can be 
well understood in the framework of the bisoliton theory of superconductivity 
for cuprates \cite{Davydov1,Davydov2,Davydov3,Davydov4,R2}. One of the  
chief results of the bisoliton model is that the potential energy of a static 
bisoliton, formed due to local deformation of the lattice, does not 
depend on the mass of the elementary lattice cell. This mass appears only 
in the kinetic energy of bisolitons. It is a paradox: 
the effect of isotope substitution on the transition temperature 
manifests itself when the electron-phonon interaction is weak, and can 
disappear when the electron-phonon interaction becomes stronger! 

The cuprates are materials with strongly correlated electrons. The Fermi-liquid 
approach is not applicable to this class of materials because the position and 
motion of each electron in these materials are correlated with those of all the 
others. If in metals, the electron-phonon coupling is week and linear, in materials 
with strongly correlated electrons, it is moderately strong and nonlinear 
\cite{R2}. Therefore, phonons interact with charge carriers in conventional 
superconductors and in the cuprates in a different manner. For example, in 
BCS-type superconductors, $T_c$ increases with lattice softening, 
while in the cuprates, $T_c$ increases with lattice stiffening, as shown 
in Fig. 2. In the BCS model of superconductivity, only acoustic phonons 
participate in electron pairing, while tunnelling studies of the spectral function 
$\alpha ^2 F( \omega )$ show that in cuprates, the charge carriers are coupled 
not only to acoustic phonons but also to optical ones \cite{R2}. 
The function $\alpha ^2 F( \omega )$ is the parameter of the electron-phonon 
interaction in the Eliashberg equations, which characterizes the coupling 
strength between charge carriers and phonon vibrations. 
The energy of these optical phonons is about 73 meV. 
In the cuprates, the 73 meV branch is associated with half-breathing-like 
oxygen phonon modes that propagate in the CuO$_2$ plane. 
\begin{figure}[t] 
\epsfxsize=0.8\columnwidth
\centerline{\epsffile{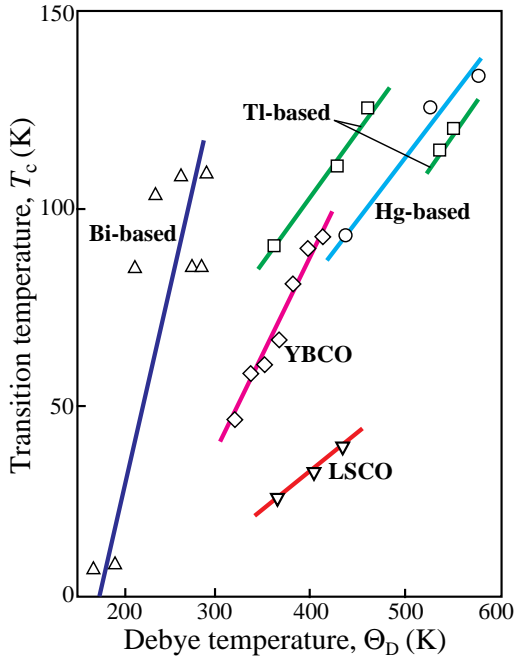}}
\caption{Critical temperature as a function of Debye temperature for
different cuprates \protect\cite{phonons}.}
\end{figure} 

The bisoliton model of superconductivity developed by Davydov for organic 
and high-$T_c$ superconductors \cite{Davydov1,Davydov2,Davydov3} is based 
on the presence of one dimensionality in the system. So, in the bisoliton theory, 
the electron transport is considered along one-dimensional chain (when the 
bisoliton model was 
created there was no knowledge about charge stripes in the cuprates). Excess 
quasiparticles present on the chain can be paired in a singlet state due to the 
interaction with local chain deformation created by them. The potential well 
formed by a short-range deformation interaction of one quasiparticle attracts 
another quasiparticle which, in turn, deepens the well. The energy of a pair of 
quasiparticles moving in the field of local deformation of the lattice with 
a velocity $v$, called a bisoliton, is given by 
\begin{equation} 
E(v) = E_p(0) + \frac{m_{bs}v^2}{2}, 
\end{equation} 
where $E_p(0)$ characterizes the position of the energy level of a static 
pair of quasiparticles beneath their Fermi level $E_F$, and 
$m_{bs}$ is the effective mass of the bisoliton. The potential energy of a 
static bisoliton $E_p(0)$ is independent of the mass of the elementary cell 
$M$. At a small density of doped charge carriers, the effective mass of a 
static bisoliton can be represented as 
\begin{equation} 
m_{bs} \simeq 2m + \frac{8 M\Delta_p}{ka^2}, 
\end{equation} 
where $m$ is the effective mass of quasiparticles; $\Delta_p$ is the pairing 
energy gap; $k$ is the coefficient of 
longitudinal elasticity, and $a$ is the distance between atoms on the chain.
The mass of a bisoliton exceeds two effective masses of quasiparticles 
forming a bisoliton. It is worth noting that the weakest point of the bisoliton 
model of superconductivity is that the Coulomb repulsion of electrons is not 
taking into account. In spite of this fact, the bisoliton theory gives pretty 
accurate values of the Cooper-pair size in hole- and electron-doped 
cuprates \cite{R2}. 

Substituting the expression for $m_{bs}$ into Eq.(1), one obtains 
\begin{equation} 
E(v) = E_p(0) + mv^2 + \frac{4 M \Delta_p v^2}{ka^2}. 
\end{equation} 
In this expression, only the last term contains the mass of the elementary 
lattice cell $M$. Therefore, in the framework of the bisoliton model, the 
isotope effect is determined by the product $\frac{\Delta_p v^2}{ka^2}$. 
Let us analyze its doping dependence. The divisor $ka^2$ is a weak function of 
$p$. The doping dependence of the pairing energy gap $\Delta_p$ in hole-doped 
cuprates is shown in Fig. 3. Figure 4 depicts the doping dependence of the 
velocity of nodal quasiparticles, determined by angle-resolved photoemission 
measurements. In Fig. 4(b), one can observe that the nodal high-energy 
quasiparticles in hole-doped cuprates move quicker in the underdoped region 
than in the overdoped region. 
Assuming that {\em nodal} quasiparticles with {\em high} energy 
play an important role in the occurrence of superconductivity in the cuprates, 
we obtain that, in the framework of the bisoliton model, the product 
$\Delta_p v_{HE}^2$ which determines the isotope effect has a doping 
dependence similar to that of $\alpha_0$ shown in Fig. 1. Thus, in the 
framework of the bisoliton theory, the doping dependence of $\alpha_0$ 
in Fig. 1 is natural. We are not going to discuss here the $\frac{1}{8}$ point in 
LSCO, as well as the weak uprise of $\alpha_0(p)$ in the overdoped region 
which can be due to several effects. 

\begin{figure}[t]
\epsfxsize=1.0\columnwidth
\centerline{\epsffile{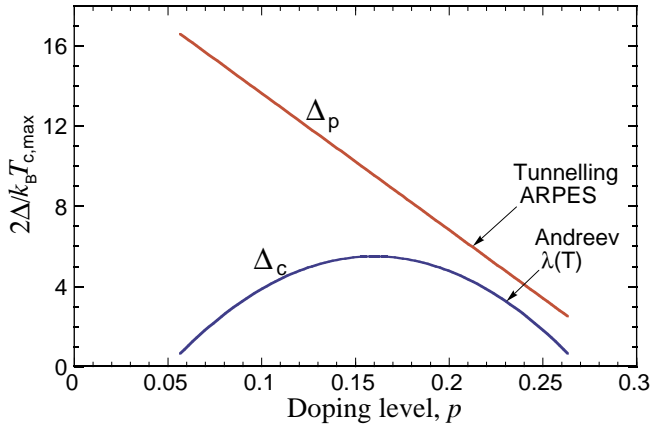}}
\caption{The phase diagram of hole-doped cuprates \protect\cite{Guy,R2}. 
$\Delta_c$ is the phase-coherence energy gap, and $\Delta_p$ is the pairing 
energy gap.}
\end{figure} 

\begin{figure}[t]
\epsfxsize=0.7\columnwidth
\centerline{\epsffile{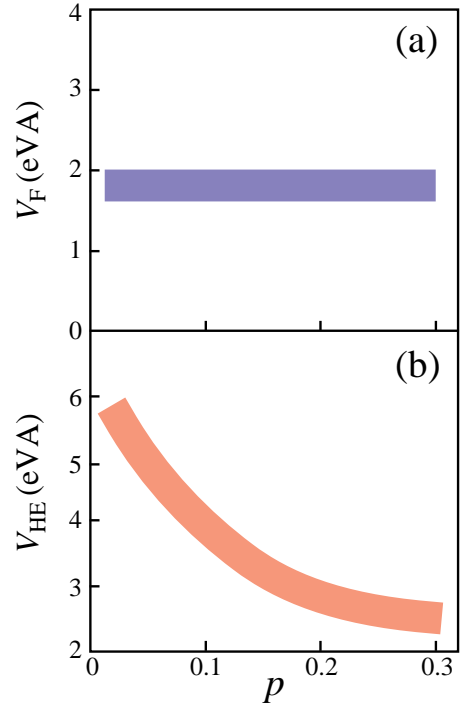}}
\caption{Velocities of nodal quasiparticles as a function of doping level $p$ in 
hole-doped cuprates \protect\cite{ARPES}. (a) The low-energy ($<$ 50 meV) 
Fermi velocity, and (b) the high-energy ($>$ 50 meV) velocity.}
\end{figure} 

To conclude, we showed here that the isotope effect in the cuprates can 
be qualitatively understood in the framework of the bisoliton theory of 
superconductivity \cite{Davydov1,Davydov2,Davydov3,Davydov4,R2,R}. 
It follows from the fact that the potential energy of a static bisoliton, formed 
due to local deformation of the lattice, does not depend on the mass of the 
elementary lattice cell. In the framework of the bisoliton model, this 
mass appears only in the kinetic energy of bisolitons.


\begin{references} 

\bibitem{BCS} Bardeen J, Cooper L N and Schrieffer J R 1957 {\it Phys. Rev.}
{\bf 108} 1175  

\bibitem{6} Maxwell E 1950 {\it Phys. Rev.} {\bf 78} 477 

\bibitem{7} Reynolds C A, 
Serin B, Wright W H and Nesbitt L B 1950 {\it Phys. Rev.} {\bf 78} 487 

\bibitem{Muller} Bednorz J G and M\"{u}ller K A 1986 {\it Z. Phys.} B {\bf 64} 
189 

\bibitem{R} Mourachkine A 2004 {\it Room-Temperature Superconductivity}, 
(Cambridge: Cambridge International Science Publishing) 

\bibitem{isotope} See {\it e g} Alexandrov A S 1992 {\it Phys. Rev.} B 
{\bf 46}  14932R, and Greco A and Zeyher R 1999 {\it Phys. Rev.} B 
{\bf 60} 1296 

\bibitem{Davydov1} Davydov A S 1988 {\it Solitons in Molecular Systems} 
(Kiev: Naukova Dumka) in Russian

\bibitem{Davydov2} Davydov A S 1990 {\it Phys. Rep.} {\bf 190} 191 

\bibitem{Davydov3} Davydov A S 1991 {\it Solitons in Molecular Systems} 
(Dordrecht: Kluwer Academic) 

\bibitem{Davydov4} Davydov A S and Kruchinin S P 1991 {\it Physica} C 
{\bf 179} 461 

\bibitem{R2} Mourachkine A 2002 {\it High-Temperature Superconductivity 
in Cuprates: The Nonlinear Mechanism and Tunneling Measurements}, 
(Dordrecht: Kluwer Academic) 

\bibitem{phonons} Ledbetter H 1994 {\it Physica} C {\bf 235-240} 1325 

\bibitem{Guy} Deutscher G 1999 {\it Nature} {\bf 397} 410 

\bibitem{ARPES} 
Zhou X J, Yoshida T, Lanzara A, Bogdanov P V, Kellar S A, Shen K M, 
Yang W L, Ronning F, Sasagawa T, Kakeshita T, Noda T, Eisaki H, Ushida S, 
Lin C T, Zhou F, Xiong J W, Ti W X, Zhao Z X, Fujimori A, Hussain Z and 
Shen Z-X 2003 {\it Nature} {\bf 423} 398 


\end{references}
\end{document}